\documentclass{article}
\usepackage{amsmath,graphicx,mlspconf}
\usepackage{xcolor}
\usepackage{cite}
\usepackage{amsmath,amssymb,amsfonts}
\usepackage{algorithmic}
\usepackage{graphicx}
\usepackage{booktabs}
\usepackage{textcomp}
\usepackage{algorithm}
\usepackage{xurl}
\usepackage[hidelinks]{hyperref}
\usepackage{placeins}
\usepackage{makecell}

\toappear{Preprint version. Accepted at 2026 IEEE International Workshop on Machine Learning for Signal Processing.}

\title{Small Language Model enabled Autonomous agent for Language-Conditioned Cognitive Radar}

\name{%
   Minhaj Uddin Ahmad$^{\dagger}$%
   \qquad Zakia Zaman$^{\dagger}$%
   \qquad Shunqiao Sun$^{\ddagger}$%
   \qquad Mizanur Rahman$^{\dagger}$
}
\address{%
   $^{\dagger}$ Department of Civil, Construction and Environmental Engineering, University of Alabama \\%
   $^{\ddagger}$ Department of Electrical and Computer Engineering, University of Alabama%
}

\begin{document}
%\ninept

\maketitle

\begin{abstract}
Modern radar systems require adapting their processing strategies in response to changing interference, clutter, and data availability. This paper introduces a framework for a small language model (SLM)-driven autonomous agent designed for language-conditioned cognitive radar, functioning as an intelligent controller for a suite of array signal processing tools. Given a natural-language command, the agent extracts radar-operation-related cues, selects an appropriate sequence of signal-processing methods, configures parameters, and invokes executable tools for numerical computation. Experiments with a synthetic uniform linear array (ULA) radar demonstrate that, given a natural-language command, the agent performs meaningful algorithm selection across diverse scenarios for sidelobe control, jammer suppression, multiple-null beamforming, coherent-source handling, and low-snapshot direction-of-arrival (DOA) estimation. Ablation results show that radar-specific prompting and physics-grounded tool execution are both required for reliable decisions and hallucination-free numerical results.
\end{abstract}

\newcommand{\cem}[1]{\textcolor{blue}{cem: #1}}

\section{Introduction}
\label{sec:intro}

Array radar systems are fundamental sensing platforms for surveillance, autonomous vehicles, robotics, weather monitoring, and spectrum-aware sensing. Their performance depends critically on the selection of an appropriate signal-processing strategy for the operating condition. For example, sidelobe-controlled conventional beamforming is preferred in cluttered scenes; adaptive beamforming is required under strong interference or jamming; and subspace or sparse direction-of-arrival (DOA) estimators are needed when targets are closely spaced, or only a small number of snapshots is available. Conventional radar-processing pipelines often encode such choices through fixed rules, lookup tables, or operator expertise~\cite{van2002optimum}. Although reliable in well-specified scenarios, these designs do not generalize well to high-level task descriptions or dynamic environments.

Cognitive radar provides a natural framework for addressing this limitation. Haykin formulated cognitive radar as an intelligent closed-loop system that learns from interactions with the environment and adapts its sensing and processing strategy accordingly~\cite{cognitive_radar_SPM_2006}. Subsequent cognitive radar studies have developed Bayesian decision-theoretic methods, Markov decision processes, fuzzy logic, reinforcement learning, waveform adaptation, spectrum sharing, and radar resource management. A recent survey by Gurbuz et al.~\cite{gurbuz2020overview} summarizes this line of work and emphasizes the perception-action-cycle view of cognitive radar. More recently, Liu et al.~\cite{cognitive_radar_2026} reviewed advanced cognitive radar principles and practical systems, highlighting closed-loop architectures that integrate environmental perception, waveform optimization, cognitive transmission, echo processing, and performance evaluation for clutter suppression and jamming countermeasures. While these systems highlight the importance of feedback-driven radar adaptation, they mainly target waveform or system-parameter optimization.

% In contrast, the problem studied here is not waveform optimization alone, but language-conditioned radar-processing workflow: given a high-level user command, the system must infer the relevant radar context, choose a valid signal-processing method, set appropriate parameters, execute the corresponding numerical tool, and report only tool-grounded quantitative results. This problem is different from generic tool use because radar-processing choices are constrained by factors such as array aperture, snapshot support, source coherence, and interference type.

\begin{figure*}[!t]
\centering
\includegraphics[width=0.99\textwidth]{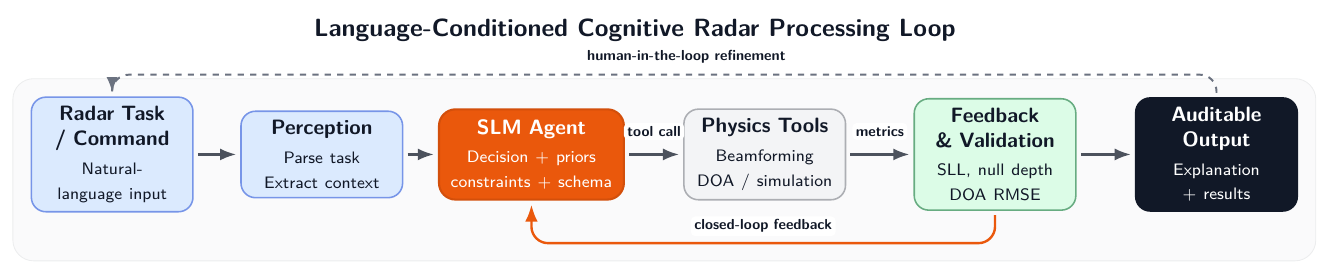}
\vspace{-5mm}
\caption{Framework for language-conditioned cognitive radar processing loop.}
\label{fig:slm_cognitive_radar_loop}
\vspace{-5mm}
\end{figure*}

Motivated by the cognitive radar paradigm, this paper presents an SLM-driven autonomous agent for language-conditioned cognitive radar, serving as an intelligent user-facing controller for array signal processing tools. Recent progress in large language models and tool-augmented agents makes this framework feasible. Brown et al.~\cite{brown2020language} demonstrated that large language models can exhibit few-shot generalization through in-context learning, and Wei et al.~\cite{wei2022chain} showed that chain-of-thought prompting can improve multi-step reasoning. Yao et al.~\cite{yao2022react} introduced the ReAct framework, which interleaves reasoning with tool invocations to ground language model decisions in external observations. Recent open-weight models, including Gemma~\cite{team2024gemma} and Qwen~\cite{yang2025qwen3}, further enable local, privacy-preserving, and potentially edge-deployable language-model agents. Radar applications impose requirements beyond generic agents: actions must be physically valid, tool parameters must meet array-processing constraints, and numerical claims must be derived from computed observations, not hallucinated values.

The main contributions of this paper are as follows: (i) We formulate language-conditioned cognitive radar as a tool-augmented SLM-agent problem, where the agent maps natural-language commands to array signal-processing actions; (ii) We implement a modular physics-grounded radar toolchain, enabling all numerical claims to be derived from executable signal-processing routines rather than model-generated arithmetic; (iii) We design radar-specific system prompt, parameter validation, state management, and transcript logging mechanisms that enforce physical constraints, reduce hallucinated numerical reporting, and make each agent decision auditable; and (iv) We evaluate the system on an expanded natural-language benchmark using both agent-level decision metrics and physics-grounded radar-performance metrics.

\vspace{-4mm}
\section{Method}
\label{sec:method}

The SLM-based autonomous agent for cognitive radar is designed using a modular architecture. Fig.~\ref{fig:slm_cognitive_radar_loop} presents the language-conditioned cognitive radar processing loop. The loop begins by parsing the input into structured radar-relevant cues, such as target direction, interference direction, clutter conditions, snapshot support, and resolution constraints. Using these cues, the SLM agent selects an appropriate processing action from a predefined library of array signal-processing methods, including conventional and adaptive beamforming, subspace-based DOA estimation, and sparse-recovery-based parameter estimation. The selected physics-based module then outputs quantitative radar metrics, such as sidelobe levels, null depths, angular estimates, and computational runtime. These metrics are  returned to the agent for validation, explanation, and the generation of the final response.

%A dedicated physics tools module provides radar-relevant signal processing algorithms. The SLM agent module then leverages the reasoning capabilities and algorithmic tools of the physics tools module to perform tasks within a closed feedback loop. The agent’s reasoning process and its interactions with the user are recorded, and all generated artifacts (such as plots) are stored to enable future auditing.

\subsection{Physics Tools}
\label{ssec:physics_engine}
The Physics Tools module is a standalone Python package built with NumPy and SciPy, which implements well-established array signal processing algorithms. The module includes three groups of tools: an array simulator, beamforming tools, and DOA estimation tools. By design, these functions provide a physics-informed foundation for the numerical outputs generated by the SLM, thereby helping to reduce hallucinations.

An array simulator generates narrowband uniform linear array (ULA) snapshot data based on a standard signal model. This simulator serves as a surrogate for actual radar hardware in our experiments. Different types of windowing weights can be applied to the data depending on the scenario and user-provided requirements, such as uniform, Dolph-Chebyshev, and Taylor. Uniform weighting gives the narrowest possible mainlobe for a given array size, while Dolph-Chebyshev and Taylor weighting provide specific sidelobe-control characteristics.

The beamforming tools include Delay-and-Sum (DAS), Minimum Variance Distortionless Response (MVDR), Minimum Power Distortionless Response (MPDR), and Linearly Constrained Minimum Variance (LCMV) beamformers. DAS is the most conventional beamforming method and applies uniform or tapered weights based on the steering vector. MVDR and MPDR minimize output interference-plus-noise power while preserving unit gain in the target direction, with the primary difference being the covariance matrix used in each method. Diagonal loading is applied to both adaptive beamformers to improve conditioning under limited snapshots. LCMV generalizes MVDR by allowing multiple  linear constraints, making it useful when nulls must be placed toward high-power jammers at known locations.

The DOA estimation tools include Multiple Signal Classification (MUSIC), Estimation of Signal Parameters via Rotational Invariance Techniques (ESPRIT), and Orthogonal Matching Pursuit (OMP). MUSIC is a subspace-based DOA estimation algorithm that uses the orthogonality between steering vectors and the noise subspace to identify source directions. ESPRIT partitions the array into two overlapping subarrays and uses rotational invariance of the signal subspace to obtain DOAs without grid search. OMP formulates DOA estimation as sparse recovery over an overcomplete angular dictionary and is useful when only limited snapshots are available.

\vspace{-3mm}
\subsection{SLM based Autonomous Agent}
\label{ssec:llm_reasoning_agent}

The reasoning engine uses Qwen3.5, a 4-bit quantized 9B-parameter model served locally via Ollama. It supports chain-of-thought reasoning and tool calls, making it suitable for agentic workflows. Classified as an SLM due to its smaller size, the 4-bit quantized model is compact enough for embedded platforms that process radar signals, such as Nvidia Jetson devices. Local, offline deployment mitigates data-exfiltration risks in classified or privacy-sensitive radar applications. Ollama’s stable API also allows future model upgrades without changing the radar signal processing pipeline. 

\subsubsection{Agentic Workflow}
\label{sssec:agentic_workflow}

The agent follows a structured \emph{observe-reason-act} loop based on the ReAct paradigm~\cite{yao2022react}. On each turn, the Ollama client sends the full message history, system prompt, user query, and all prior assistant turns with their tool results to the model and streams the response. The response may contain (a) reasoning, (b) tool calls in the provider's function-call schema, or (c) both. The loop ends when the model produces a turn without tool calls, which is treated as the final human-readable answer. The agent never performs arithmetic directly: all numerical computations (e.g., covariance estimation, weight calculation, subspace decomposition) are delegated to physics tools. The model only decides which tool to call with which parameters and then synthesizes the tool results into an expert explanation.

\subsubsection{System Prompt Engineering}
\label{sssec:sys_prompt}

The system prompt is the primary knowledge-encoding mechanism in this study. Because Qwen~3.5 is a general-purpose model with no radar-specific fine-tuning, the system prompt must supply all domain knowledge required for correct algorithm selection. The system prompt has five parts:

1. Opening sentences define the agent as ``an expert radar array signal processing agent'' and state key rules: assess algorithmic suitability, avoid manual numeric computation, and always use tools for signal processing. Each tool or algorithm has a one-line description of its operating conditions (e.g., MUSIC: ``Needs L $\gg$ N snapshots; fails on coherent sources unless spatial smoothing applied."), which encodes the signal processing knowledge the agent uses to choose tools.

2. The prompt embeds radar array related knowledge. For example: ``If jammer or high-power interference → simulate\_ula with the jammer modeled. If a target signal-free training period is plausible, set signal\_of\_interest\_index and use beamform\_mvdr; otherwise beamform\_mpdr is more suitable." These priors supply reliable expert rules that a small model might not recall, keeping behavior stable even if the SLM changes.

3. Inviolable constraints (e.g., Rayleigh resolution limit, N-1 source capacity, MUSIC rank conditions, minimum sidelobe level without tapering) are specified to block physically impossible claims or parameter choices.

4. The final part requires that every tool call be preceded by at least one sentence of explicit reasoning describing observations, the chosen tool, and the rationale. It also requires a non-empty final answer that reports numerical results only from tool outputs. The agent loop enforces these rules by inserting a “nudge” message if the model ends with an empty answer.

\subsubsection{State Management}
\label{sssec:state_management}

The tool calls are stateless from the model's perspective. Each call receives only the arguments the model supplies. Yet the radar processing loop here has a natural sequential structure: simulate, then run signal processing, then analyze, so the handlers share a single \texttt{RunContext} ``data-class'' that persists for the duration of an invocation. When a tool call returns a result, the numerical values are stored in that RunContext. All subsequent tool calls can use these values and update as needed. This design cleanly separates the model's stateless reasoning from stateful physics-grounded numerical values.

\subsubsection{Agentic Loop}
\label{sssec:agentic_loop}

The core of the integration is the \texttt{SLMAgent.run()} loop (Algorithm~\ref{alg:loop}). It initializes the message history with the system prompt and user query, then runs for up to \texttt{max\_steps = 8} iterations. On each iteration, the Ollama client is called with the full history and tool schema list. If the model returns tool calls, their handlers are invoked, the results are appended as \texttt{role:~"tool"} messages, and the loop continues. A key drawback of SLMs in our study is premature conversation termination, especially with long context windows. This violates the system prompt, which requires a final answer quoting numerical results from tool calls, so the agent loop adds a recovery step. When the radar processing loop detects early termination with no tool calls and empty content, it does not treat this as the end of the conversation. Instead, it injects a short message reminding the model to provide a concrete response and summarize the tool call results so far. This is appended as a \texttt{role:"user"} turn, and the model is called again. This nudging mechanism successfully recovers from premature termination events.

\begin{algorithm}
\caption{SLMAgent.run()}
\label{alg:loop}
\begin{algorithmic}
\STATE transcript $\leftarrow$ [\{system\}, \{user: query\}]
\FOR{step $= 0, 1, \ldots, \texttt{max\_steps}$}
    \STATE content, tool\_calls $\leftarrow$ stream\_chat(transcript)
    \IF{tool\_calls is empty and content is empty}
        \STATE inject nudge message; retry once; else break
    \ENDIF
    \STATE append \{assistant: content, tool\_calls\} to transcript
    \IF{tool\_calls is empty}
        \STATE \textbf{break} \COMMENT{final answer turn}
    \ENDIF
    \FOR{each tc in tool\_calls}
        \STATE result $\leftarrow$ registry[tc.name](**tc.args)
        \STATE append \{tool: result\} to transcript
    \ENDFOR
\ENDFOR
\RETURN transcript
\end{algorithmic}
\end{algorithm}

\subsubsection{Output Artifacts}
\label{sssec:output_artifacts}

Each run creates a timestamped directory containing two artifacts. A transcript file contains a complete decision trace: every system, user, assistant, and tool message in order, allowing future audit of the agent's decisions and the exact parameters passed to each physics call. A plots directory stores beampattern and pseudo-spectrum figures. The command-line interface (CLI) displays the agent's decisions and tool calls in real time in the terminal, giving the operator continuous visibility into the agent's decision-making process.

\subsubsection{Configuration and Extensibility}
\label{sssec:configuration}

Array defaults, model identity, Ollama host, generation parameters, angular grid resolution, and output figure resolution are specified in \texttt{config.yaml}, making the system straightforwardly reconfigurable without code changes. Swapping the underlying LLM requires only changing \texttt{model.name}; any Ollama-hosted model that advertises \texttt{tools} capability in its manifest is compatible. Adding a new algorithm follows a three-step pattern: implement the math in the physics module, add a JSON schema entry to \texttt{TOOL\_SPECS}, and write a handler in \texttt{build\_tool\_registry()}. The system prompt should then be updated with a one-sentence description of when to prefer the new tool.

% ============================================================
\section{Experimental Evaluation}
\label{sec:result}
% ============================================================

This section evaluates the presented SLM-based cognitive radar agent from two complementary perspectives. 
First, we evaluate the \emph{agentic decision-making capability}: whether the SLM selects the correct radar signal-processing tool, produces valid tool calls, uses physically meaningful parameters, and avoids hallucinating numerical results. Second, we evaluate the \emph{physics-grounded radar performance}: whether the selected tools achieve the requested sidelobe level, jammer suppression, and DOA estimation accuracy. In addition, we inspect specific failure cases in Table~\ref{tab:failure_modes} to identify the root cause, which will guide future improvements. Unless otherwise stated, the default ULA contains $N=16$ elements with inter-element spacing $d=0.5\lambda$. The default number of snapshots is $L=200$ and the default SNR is $10$~dB.

We construct a benchmark of natural-language radar-processing tasks across six representative categories, as described in Table ~\ref{tab:benchmark_tasks}. For each category, we create multiple paraphrased user commands so that the evaluation tests natural-language robustness rather than memorization of a fixed prompt template. Each prompt is executed multiple times to account for stochastic variation in SLM responses. For each trial, an expert-defined reference label specifies the acceptable tool or tool sequence. 

\begin{table}[t]
\centering
\caption{Natural-language benchmark tasks categories.}
\label{tab:benchmark_tasks}
\footnotesize
\resizebox{\linewidth}{!}{
\begin{tabular}{p{0.22\linewidth} p{0.40\linewidth} p{0.40\linewidth}}
\toprule
\textbf{Task category} & \textbf{Example user command} & \textbf{Expected tool choice} \\
\midrule
Sidelobe control 
& ``Scan toward $30^\circ$ with sidelobes below $-30$~dB.'' 
& DAS with Chebyshev/Taylor taper. \\

Single jammer 
& ``Track a target at $10^\circ$; a strong jammer appears at $-20^\circ$.'' 
& MVDR if training data are available; MPDR otherwise \\

Multiple jammers 
& ``Track the target at $10^\circ$ and suppress jammers at $-30^\circ$ and $25^\circ$.'' 
& LCMV with one distortionless constraint and two null constraints \\

Low-snapshot DOA 
& ``Only 10 snapshots are available; estimate two closely spaced DOAs.'' 
& ESPRIT or OMP; avoid conventional MUSIC when $L$ is too small \\

Coherent-source DOA 
& ``Two coherent sources are present; estimate their DOAs.'' 
& MUSIC with forward spatial smoothing or an appropriate sparse method \\

Underspecified prompt 
& ``There is strong interference; choose a robust processing strategy.'' 
& Ask for missing information or make explicit, physically reasonable assumptions \\
\bottomrule
\end{tabular}}
\end{table}

% \begin{table*}[htbp]
% \centering
% \caption{Agent-level quantitative evaluation.}
% \label{tab:agent_ablation}
% \resizebox{0.9\linewidth}{!}{
% \begin{tabular}{lcccccc}
% \toprule
% \textbf{Method} 
% & \textbf{Alg. acc.} 
% & \textbf{Tool valid} 
% & \textbf{Param. valid} 
% & \textbf{Halluc.-free} 
% & \textbf{Radar Metric}
% & \textbf{E2E success} \\
% \midrule

% Rule-based controller & $15.0\%$ (9/60) & $20.0\%$ (12/60) & $16.7\%$ (10/60) & $100.0\%$ (60/60) & $31.7\%$ (19/60) & $15.0\%$ (9/60) \\
% SLM w/o radar prompt + tools & $67.8\%$ (122/180) & $96.7\%$ (174/180) & $93.3\%$ (168/180) & $90.6\%$ (163/180) & $73.9\%$ (133/180) & $52.8\%$ (95/180) \\
% SLM w/ radar prompt, no tools & $96.1\%$ (173/180) & -- & -- & $52.8\%$ (95/180) & -- & $52.2\%$ (94/180) \\
% Full SLM agent & $91.1\%$ (164/180) & $100.0\%$ (180/180) & $98.3\%$ (177/180) & $98.9\%$ (178/180) & $88.3\%$ (159/180) & $86.1\%$ (155/180) \\
% \bottomrule
% \end{tabular}}
% \end{table*}

\begin{table*}[htbp]
\centering
\vspace{-2mm}
\caption{Agent-level quantitative evaluation. All values reported with 95\% Wilson confidence intervals.}
\label{tab:agent_ablation}
\resizebox{\linewidth}{!}{
\begin{tabular}{lcccccc}
\toprule
\textbf{Method} 
& \textbf{Alg. acc.} 
& \textbf{Tool valid} 
& \textbf{Param. valid} 
& \textbf{Halluc.-free} 
& \textbf{Radar Metric}
& \textbf{E2E success} \\
\midrule
Rule-based controller & $15.0\%$ [8.1, 26.1] & $20.0\%$ [11.9, 31.6] & $16.7\%$ [9.3, 28.0] & $100.0\%$ [94.0, 100.0] & $31.7\%$ [21.4, 44.2] & $15.0\%$ [8.1, 26.1] \\
SLM w/o radar prompt + tools & $67.8\%$ [60.6, 74.2] & $96.7\%$ [92.9, 98.5] & $93.3\%$ [88.7, 96.2] & $90.6\%$ [85.4, 94.1] & $73.9\%$ [67.1, 79.7] & $52.8\%$ [45.5, 60.0] \\
SLM w/ radar prompt, no tools & $96.1\%$ [92.1, 98.1] & -- & -- & $52.8\%$ [45.5, 60.0] & -- & $52.2\%$ [44.9, 59.4] \\
Full SLM agent & $91.1\%$ [86.1, 94.4] & $100.0\%$ [97.9, 100.0] & $98.3\%$ [95.2, 99.4] & $98.9\%$ [96.0, 99.7] & $88.3\%$ [82.7, 92.3] & $86.1\%$ [80.3, 90.4] \\
\bottomrule
\vspace{-12mm}     % reduce space after the table.
\end{tabular}}
\end{table*}

% ------------------------------------------------------------
\subsection{Agent Decision Accuracy and Ablation Study}
\label{ssec:agent_ablation}
% ------------------------------------------------------------

Table~\ref{tab:agent_ablation} compares the proposed full agent against three ablated baselines. The first baseline is the SLM agent without the radar-specific system prompt, while retaining tool access. The second baseline keeps the radar-specific prompt but disables tool execution, forcing the SLM to answer without physics-grounded computation. The third baseline is a fixed, rule-based controller that uses keyword matching. This comparison separates the contribution of domain-specific prompting, tools, and natural-language reasoning. In total, we evaluate 180 trials, comprising six task categories, each with ten paraphrased prompts, each run with three different seeds. 

We report the following agent-level metrics: {(i) Algorithm-selection accuracy}: percentage of trials in which the selected tool or tool sequence matches the expert-defined acceptable set; {(ii) Valid-tool-call rate}: percentage of trials in which the SLM produces valid tool calls with valid arguments; {(iii) Parameter-validity rate}: percentage of trials in which the selected parameters satisfy physical and implementation constraints, such as $K<N$, valid angle ranges, and feasible sidelobe specifications; {(iv) Hallucination-free rate}: percentage of trials in which the final answer quotes only numerical values returned by the tools; {(v) Radar metric}: percentage of results returned by the SLM agent is numerically accurate (e.g., the DOA estimation angle is the same as the ground truth); and {(vi) End-to-end success rate}: percentage of trials that satisfy both the agent-level decision criteria and the radar-performance criteria. Each reported percentage in Table~\ref{tab:agent_ablation} is computed over these 180  trials.

The ablation study is designed to test the contributions of physics tools and the system prompt. Without the radar-specific system prompt, the SLM often lacks the domain priors required to distinguish between superficially similar methods, such as MVDR and MPDR. Without tool execution, the SLM can still describe a plausible algorithm, but cannot provide reliable numerical values such as null depth, sidelobe level, or DOA RMSE. The rule-based controller is reliable for simple keyword-matched commands but is less robust to underspecified or paraphrased user requests.

% ------------------------------------------------------------
\subsection{Beamforming and Sidelobe-Control Results}
\label{ssec:beamforming_results}
% ------------------------------------------------------------

We first evaluate whether the agent can translate natural-language sidelobe specifications into appropriate beampattern-design choices. For a user request requiring a $30^\circ$ scan direction and sidelobes below $-30$~dB, the agent selected a conventional delay-and-sum beamformer with a Dolph-Chebyshev taper. 
This is the appropriate choice because the Dolph-Chebyshev taper provides the narrowest mainlobe for a prescribed equiripple sidelobe level. Table~\ref{tab:sidelobe_results} provides a comparative analysis of the uniform, Taylor, and Dolph–Chebyshev tapering schemes. The uniform taper provides the narrowest mainlobe but insufficient sidelobe suppression. The tapered designs trade mainlobe width for lower sidelobes, matching the physical design trade-off expected in clutter-limited radar operation. Figure~\ref{fig:beamforming_results}~(Left) presents beamforming results generated by the physics tools. These results show that the agent selects a meaningful taper rather than blindly applying the highest-resolution uniform beamformer.

% ------------------------------------------------------------
\subsection{Adaptive Jamming Suppression}
\label{ssec:jamming_results}
% ------------------------------------------------------------

\begin{figure}[htbp]
    \centering
    \includegraphics[width=0.98\linewidth]{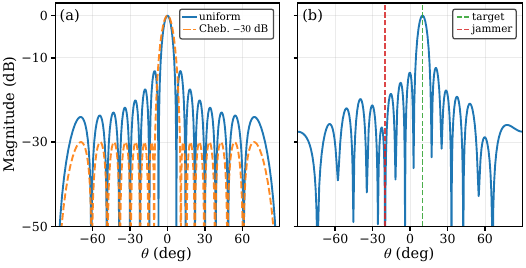}
    \vspace{-4mm}
    \caption{Beamforming results from the physics tools. 
Left: DAS beampatterns with uniform and Dolph–Chebyshev tapering, illustrating the trade-off between mainlobe width and sidelobe suppression. 
Right: adaptive beampattern for a target at $10^\circ$ and jammer at $-20^\circ$, showing a deep null at the jammer while preserving the target response.}
    \label{fig:beamforming_results}
    \vspace{-4mm}
\end{figure}

We next evaluate adaptive beamforming under high-power jamming. The user specifies a desired target direction and one or more jammer directions. The agent must select the appropriate adaptive beamformer and pass the correct target and interference parameters to the physics engine. The adaptive-jamming task is particularly useful for evaluating the agent because it requires more than keyword matching. When a signal-free training interval is available, MVDR should use an interference-plus-noise covariance estimate and preserve the target response. When such a training interval is unavailable, MPDR must operate from the full data covariance and is more vulnerable to signal self-nulling. For multiple simultaneous jammers, LCMV is the appropriate generalization because it imposes one distortionless response constraint in the target direction and multiple null constraints in the jammer directions. Table~\ref{tab:jamming_results} presents the performance under different jamming conditions, and Figure~\ref{fig:beamforming_results}~(Right) presents a plot generated by physics tools in a single jammer condition.

\begin{table}[t]
\centering
\caption{Sidelobe-control performance for beamformers.}
\label{tab:sidelobe_results}
\resizebox{\linewidth}{!}{
\begin{tabular}{lcccc}
\toprule
\textbf{Taper} & \textbf{Steer angle} & \textbf{Requested SLL} & \textbf{Achieved SLL} & \textbf{3-dB BW} \\
\midrule
Uniform & $0^\circ$ & -- & $-13.15$~dB & $6.348^\circ$ \\
Dolph-Chebyshev & $0^\circ$ & $-30$~dB & $-30.00$~dB & $7.966^\circ$ \\
Taylor & $0^\circ$ & $-30$~dB & $-30.06$~dB & $8.054^\circ$ \\
Dolph-Chebyshev & $30^\circ$ & $-30$~dB & $-30.00$~dB & $9.212^\circ$ \\
Dolph-Chebyshev & $30^\circ$ & $-40$~dB & $-40.00$~dB & $10.384^\circ$ \\
\bottomrule
\end{tabular}}
\end{table}

% ------------------------------------------------------------
\subsection{DOA Estimation Under Constraints}
\label{ssec:doa_results}
% ------------------------------------------------------------

The third set of experiments evaluates whether the agent selects suitable DOA estimators under limited snapshots, closely spaced sources, and coherent-source conditions. 
This setting is important because different DOA methods have different failure modes. 
MUSIC requires a reliable covariance estimate and a valid noise subspace, ESPRIT is often more stable with limited snapshots, and OMP is suitable when the data record is extremely short or sparse recovery is preferred. Table~\ref{tab:doa_results} reports representative DOA-estimation results from the physics tool calls. A representative example, in the low-snapshot experiment, the agent does not simply default to a single algorithm for all DOA-estimation tasks. Instead, it recognizes that $L$ is small relative to $N$ and selects ESPRIT, which is a stable choice when the sample covariance is poorly conditioned. 

\begin{table}[t]
\centering
\caption{Adaptive beamforming in jamming conditions.}
\label{tab:jamming_results}
\resizebox{\linewidth}{!}{
\begin{tabular}{lccccc}
\toprule
\textbf{Scenario} 
& \textbf{Selected tool} 
& \textbf{Target} 
& \textbf{Jammer(s)} 
& \textbf{Null depth} 
& \textbf{Target gain} \\
\midrule

Single jammer, training available & MVDR & $10^\circ$ & $-20^\circ$ & $-81.77$~dB & $0.00$~dB \\
Single jammer, no training & MPDR & $10^\circ$ & $-20^\circ$ & $-55.29$~dB & $-0.01$~dB \\
Two jammers & LCMV & $10^\circ$ & $-30^\circ,+25^\circ$ & $-120.00$~dB & $0.00$~dB \\
Steering mismatch & MVDR (Loaded) & $10^\circ$ & $-20^\circ$ & $-85.52$~dB & $0.00$~dB \\

\bottomrule
\end{tabular}}
\end{table}

\begin{table}[t]
\centering
\caption{DOA-estimation results under limited snapshots and closely spaced sources.}
\label{tab:doa_results}
\resizebox{\linewidth}{!}{
\begin{tabular}{cccccc}
\toprule
\textbf{$N$} 
& \textbf{$L$} 
& \textbf{SNR} 
& \textbf{True DOAs} 
& \textbf{Selected tool} 
& \textbf{Estimated DOAs / RMSE} \\
\midrule
$8$ & $10$ & $10$~dB & $[10^\circ,15^\circ]$ & ESPRIT & $[9.570^\circ,14.876^\circ]$ / RMSE $=0.316^\circ$ \\
$16$ & $10$ & $10$~dB & $[30^\circ,32^\circ]$ & ESPRIT & $[29.794^\circ,32.764^\circ]$ / RMSE $=0.559^\circ$ \\
$16$ & $1$ & $10$~dB & $[10^\circ,15^\circ]$ & OMP & $[6.500^\circ,13.750^\circ]$ / RMSE $=2.628^\circ$ \\
$16$ & $200$ & $10$~dB & $[10^\circ,15^\circ]$ & MUSIC & $[10.000^\circ,15.000^\circ]$ / RMSE $=0.000^\circ$ \\
$16$ & $50$ & $0$~dB & $[10^\circ,15^\circ]$ & ESPRIT & $[10.116^\circ,15.284^\circ]$ / RMSE $=0.217^\circ$ \\
$16$ & $100$ & $10$~dB & $[10^\circ,15^\circ]$ & FSS-MUSIC & $[10.000^\circ,15.000^\circ]$ / RMSE $=0.000^\circ$ \\
\bottomrule
\end{tabular}}
\end{table}
\vspace{-5mm}

\begin{table}[htbp]
\centering
\caption{Failure modes observed in the benchmark.}
\label{tab:failure_modes}
\footnotesize
\resizebox{\linewidth}{!}{
\begin{tabular}{p{0.15\linewidth} p{0.09\linewidth} p{0.07\linewidth} p{0.80\linewidth}}

\toprule
\textbf{Scenario} & \textbf{Executed} & \textbf{Count} & \textbf{What's going on} \\
\midrule
\multicolumn{4}{l}{\textit{Algorithm mis-selection (wrong tool dispatched)}} \\
Low-snapshot DOA & \texttt{MUSIC} & 8/180 &  Model recites instruction that MUSIC needs well-conditioned $R$, but still dispatches \texttt{estimate\_doa\_music}. A better-engineered system prompt to embed the relationship between low snapshot and the ``well-condition'' of $R$ matrix may help. \\

Sidelobe shaping & \texttt{MVDR} & 4/180 & Phrasings like ``reflections from elsewhere are negligible'' are read as an adaptive-null problem; the model calls MVDR instead of selecting a low-sidelobe taper on DAS. \\

Sidelobe shaping & \texttt{LCMV} & 3/180 & Phrasings like ``junk everywhere off-axis'' are read as needing many explicit nulls; the model invents a list of null angles (e.g. \texttt{[15, 30, 45, 60, 75]}) rather than picking a tapered DAS that suppresses all sidelobes. \\
Coherent-source DOA & \texttt{ESPRIT} & 1/180 & Coherent sources break ESPRIT's rotational-invariance assumption; the system prompt routes the scenario to MUSIC with spatial smoothing, but the model dispatches ESPRIT anyway. \\

\midrule
\multicolumn{4}{l}{\textit{Right tool, wrong result (parameter or radar check fails)}} \\

Low-snapshot DOA & \texttt{OMP} & 5/180 & Correct algorithm choice, but the angular dictionary is too coarse for the closely-spaced sources; sparse recovery snaps to grid bins far from the true DOAs (RMSE $\approx 16$--$35^\circ$ vs the $2^\circ$ threshold). \\

Sidelobe shaping & \texttt{DAS} & 3/180 & Right algorithm, but the model passes a positive \texttt{sll\_db} value (prompt says $-30$~dB; model sends $+30$); the taper synthesis returns NaN and the SLL check fails on the NaN. \\

Single-jammer suppression & \texttt{MVDR} & 1/180 & Right algorithm, but the model omits \texttt{soi\_index} on \texttt{simulate\_ula}; without signal-free training, the adaptive null sits at $-17.9$~dB instead of the $-25$~dB target. \\

\bottomrule
\end{tabular}}
\end{table}

\begin{table}[!htbp]
\centering
\caption{Runtime of the local SLM-agent system.}
\label{tab:runtime_results}
\resizebox{\linewidth}{!}{
\begin{tabular}{lcccc}
\toprule
\textbf{Task} 
& \textbf{Model latency} 
& \textbf{Tool time} 
& \textbf{Total time} 
& \textbf{No. tool calls} \\
\midrule
Sidelobe control & $10.16$~s & $0.393$~s & $10.55$~s & $2$ \\
Single-jammer MVDR & $9.78$~s & $0.357$~s & $10.14$~s & $2$ \\
Multiple-jammer LCMV & $8.52$~s & $0.465$~s & $8.99$~s & $2$ \\
Low-snapshot DOA & $13.21$~s & $0.240$~s & $13.45$~s & $3$ \\
\bottomrule
\end{tabular}}
\end{table}

% ------------------------------------------------------------
\subsection{Runtime and Local Deployment Feasibility}
\label{ssec:runtime_results}
% ------------------------------------------------------------
We report the runtime of the local SLM-agent loop, including SLM generation, tool-call parsing, physics tool execution, and final response generation. Because physics computations are lightweight compared to SLM inference, local model generation latency is the main bottleneck. Table~\ref{tab:runtime_results} summarizes representative runtimes. Although the present experiments are run on a local workstation with an Intel i9-14900F CPU and 64GB of RAM, the use of a quantized 9B-parameter SLM and lightweight NumPy/SciPy physics tools suggests a feasible path toward deployment on embedded GPU platforms, such as the NVIDIA Jetson family. The current implementation is self-contained and thus privacy-preserving, which is desirable for radar applications involving sensitive operational conditions.

% ============================================================
\section{Conclusion and Future Direction} \label{sec:conclusion}
% ============================================================

This paper presents an SLM agent for radar array signal processing. The system demonstrates that a small, locally hosted language model (Qwen~3.5:9b) can serve as a reliable cognitive controller for a suite of signal processing tools, provided it is given a carefully engineered system prompt that encodes algorithm-selection priors, physical sanity checks, and behavioral-discipline constraints. Several limitations remain. The current evaluation uses synthetic ULA data and a fixed radar-tool suite, so the results do not yet demonstrate robustness on hardware radar data, more general array geometries, or complex propagation environments. The agent also depends on prompt-level priors and tool-schema constraints rather than a learned long-term adaptation mechanism. Future work will target real-world radar datasets, non-ULA and MIMO configurations, stricter consistency checks between user commands and tool-call arguments, and closed-loop waveform or sensing-policy adaptation.

% ============================================================
\section{Open Science} \label{sec:open-science}
% ============================================================

The source code of this study is available at \url{https://github.com/minhaj6/cognitive-radar-SLM}

% References should be produced using the bibtex program from suitable
% BiBTeX files (here: strings, refs, manuals). The IEEEbib.bst bibliography
% style file from IEEE produces unsorted bibliography list.
% -------------------------------------------------------------------------
\bibliographystyle{IEEEbib}
\footnotesize
\bibliography{ref}

\end{document}